\documentclass[conference]{IEEEtran}
\IEEEoverridecommandlockouts
\usepackage[nocompress]{cite}

\usepackage{amsmath,amssymb,amsfonts}
\usepackage{algorithmic}
\usepackage{graphicx}
\usepackage{textcomp}
\usepackage{xcolor}
\def\BibTeX{{\rm B\kern-.05em{\sc i\kern-.025em b}\kern-.08em
    T\kern-.1667em\lower.7ex\hbox{E}\kern-.125emX}}
\usepackage{enumitem}
\setlist[itemize]{leftmargin=*}   
\usepackage{amsmath}
\usepackage{booktabs}
\usepackage{tabularx}
\usepackage{threeparttable}
\allowdisplaybreaks

\begin{document}

\title{A Visualized Malware Detection Framework with CNN and Conditional GAN 
\thanks{This work was supported in part by the Center for Cyber-Physical Systems (C2PS), Khalifa University; and in part by the Technology Innovation Institute (TII) under Grant 8434000379}
}

\author{
\IEEEauthorblockN{Fang Wang}
\IEEEauthorblockA{\textit{C2PS, Department of Electrical}\\
\textit{Engineering and Computer Science} \\
\textit{Khalifa University}\\
Abu Dhabi, United Arab Emirates \\
florencewong@protonmail.com}
\and
\IEEEauthorblockN{Hussam Al Hamadi}
\IEEEauthorblockA{\textit{College of Engineering and IT} \\
\textit{University of Dubai}\\
Dubai, UAE \\
halhammadi@ud.ac.ae}
\and
\IEEEauthorblockN{Ernesto Damiani}
\IEEEauthorblockA{\textit{C2PS, Department of Electrical}\\
\textit{Engineering and Computer Science} \\
\textit{Khalifa University}\\
Abu Dhabi, United Arab Emirates \\
ernesto.damiani@ku.ac.ae}
}
\maketitle

\begin{abstract}
Malware visualization analysis incorporating with Machine Learning (ML) has been proven to be a promising solution for improving security defenses on different platforms. In this work, we propose an integrated framework for addressing common problems experienced by ML utilizers in developing malware detection systems. Namely, a pictorial presentation system with extensions is designed to preserve the identities of benign/malign samples by encoding each variable into binary digits and mapping them into black and white pixels. A conditional Generative Adversarial Network based model is adopted to produce synthetic images and mitigate issues of imbalance classes. Detection models architected by Convolutional Neural Networks are for validating performances while training on datasets with and without artifactual samples. Result demonstrates accuracy rates of 98.51\% and 97.26\% for these two training scenarios. 
\end{abstract}


\section{Introduction}
The rapidly increasing number of cyberattacks and data breaches over years caused by malicious software (malware) has raised serious issues. Consequently, identifying malware using machine learning techniques has become a shortcut method to keep up with the complex malware metamorphoses nowadays. The visualization analysis approach \cite{nataraj2011malware} combined with advanced Machine Learning (ML) techniques has been proved to be a promising solution for malicious software detection. The approach extracts behaviors' features, such as network activities, instruction sequences, and system calls \cite{kim2019improvement, tobiyama2016malware} to convert them into an image, thus, preserving abnormal patterns as special textures for further image classification and detection purposes. An advantage of malware visualization analysis is that it enhances recognizabilities for both human and ML classifiers, regardless of other information such as technique modifications.

Extensive research on malware visualization analysis has been bridged with various ML methods and made progress in tackling different sets of problems.  For example, the choice between grayscale or colored images as transformed outputs is being explored and debated for achieving better results \cite{hsien2018, vu2019convolutional, nataraj2011malware} and lowering computational costs. Moreover, Convolutional Neural Networks (CNNs) with different architectures \cite{gibert2019using, vasan2020image, xiao2020malfcs} have been experimented with to improve the accuracy of detection performance and malware classifications. Additionally, the class imbalance problem is common due to uneven distributions of benign software and varieties of malware.  Several models were proposed to handle such problems, including a CNN based model with weighted softmax loss \cite{yue2017imbalanced}, a DenseNet model with reweighting losses method \cite{hemalatha2021efficient}, an ensemble learning model \cite{zhang2016using}, and others. Meanwhile, Generative Adversarial Network (GAN) \cite{goodfellow2014generative} has led to a significant foresight by generating images that resemble the original data, promising a better solution for balancing data. Reference \cite{kim2018zero} employed transferred deep convolutional GAN (tDCGAN) and generated fake malware for simulating zero-day attacks. In \cite{hu2017generating} work, adversarial malware examples were modeled to improve detection on black-box attacks. However, most previous studies have focused on very specific problems.

In this work, we propose an integrated framework that involves a Pictorial Representation System (PRS) with applicable features under various scenarios, a conditional Generative Adversarial Network (cGAN) based image augmentation model for mitigating data scarcity, and several CNN-based detection models for performance comparison on different datasets.  The proposed model was practically assessed using TensorFlow \cite{tensorflow2015-whitepaper} and Keras \cite{chollet2015keras} Python library.

The contribution of this work is as follows:
\begin{itemize}
\item We developed a new visualized system that targets behavior patterns of malware with a flexible mapping algorithm and could be implemented in different situations. Compared to previous visualization methods, the simplicity of only two RGB color codes - $(255, 255, 255)$ and $(0, 0, 0)$ - involved in the transformed image significantly reduces the complexity of CNN architecture and computational costs. The images can be easily identified by both human and programmed classifiers.
\item The synthetic augmentation of malware images method on the basis of cGAN allows for strengthening the robustness of malware detection solutions. The conditional parameter setup in GAN model contributes to effective image generation with targeted classes. The approach is particularly suitable for balancing malware pictorial datasets.
\item Without the need for complex feature engineering tasks on CNNs, the proposed malware detection model achieves higher accuracy rates of $98.51\%$ and $97.26\%$ while training on original and artificial sets. The model has high computational performance, achieving an efficient malware detection system and demonstrating a successful visualization system and an image augmentation method.
\end{itemize}

The rest of this paper is organized as follows; Section 2 provides a thorough explanation of our proposed framework. Results from the experiment and comparison are described in Section 3. The conclusion of our contributions and our suggestions for further work are provided in Section 4. 

\section{Proposed Framework}
Our research approach takes four steps to realize the application of a detection framework on pictorial formatted data, namely tabular data preparation and augmentation, pictorial representation system development, and, CNN and cGAN modelling.  First,  we implement Synthetic Minority Oversampling Technique (SMOTE) \cite{chawla2002smote} on tabular data to synthesize new examples for the minority class (malign) and to obtain a balanced training dataset as baseline training set.  Meanwhile, we design a binary pictorial system to transform each tabular data sample from training and testing datasets into a $64\times64$ pixels grayscale image. Second, a baseline CNN model is developed for image classification and also for comparing detection performance while excluding and including artificial pictorial data. Third, a cGAN model is designed to generate artificial pictorial data, which will be included into the original imbalanced training dataset for new balancing purpose. Finally, the CNN model is tuned to obtain new result based on the artificial dataset. 
Through this procedure, we will be able to learn how the detection model performs with the aid of pictorial data transformation. Additionally, we expect that the visualization approach makes malign samples easily distinguishable from the benign samples.  Due to the information loss caused by the pictorial transformation, we expect a slightly lower accuracy while including artificial image data in the training set.
\subsection{Tabular Data Preparation and Augmentation}
\subsubsection{Preparation}
1693 malign samples are firstly extracted from 50 malware that are validated by VirusTotal and collected from the Androzoo \cite{allix2016androzoo} database. Within a sandbox environment, each of these 50 malware was observed in action on an Android-based smartphone to gather the malign samples. To avoid the biases related to data coming from different sources, 60 trusted apps from Google Play Store and factory/system are selected as the source providing 3228 benign samples. In total 1465 malign samples and 3000 benign samples are randomly selected as our training dataset. The remain parts of data belong to the testing dataset - a balanced set with the same amount (228) of malign and benign samples. Applications' behaviors across the kernel's system were collected from a probe we developed called map. The final datasets include 128 variables representing 128 time windows of same duration, in which the number of system call (behavior) is recorded. We also label the malign sample as 1 and benign as 0.
\subsubsection{Augmentation - SMOTE}
Comparing with the data from benign apps, data from malware are limited and fall into minority category. The training dataset suffers from an imbalanced issue with 3000 data in majority class and only 1465 belonging to the minority class. Therefore, SMOTE is implemented to balance the training dataset that will be used to the CNN baseline model.  Instead over-sampling with replacement of data in the minority class, the approach of SMOTE is to over-sample data by creating ``synthetic'' examples. Namely, for each sample $t_{j} $ in minority class, $k$ nearest neighbors of this sample will be searched. Then a random neighbor will be chosen and defined as $t^{'}$.  Number $\alpha, \alpha \in [0,1]$,  is randomly generated. Then, as a new artificial sample,  $ t_\mathrm{new} $ is introduced by:
\begin{equation}
t_\mathrm{new}  = t_{j} + (t^{\prime} - t_{j}) \alpha \label{eq1}
\end{equation}
By implementing SMOTE method, the training dataset is balanced with 3000 samples in each category.  
\subsection{Pictorial Representation System (PRS)}
\subsubsection{PRS Denotation}
A pictorial representation system is designed to transfer tabular data into image data. We propose $J = 128$ as a fundamental size for PRS, where $J$ is the number of variables, and use ``padding'' for handling other sizes of $J$. The output image is set to square, with each dimension $D = J/2$, and $K$ is the number of pixels for each variable on the output image as well as the upper limit of the number of binary digits. We set $K = J/4$.  In our case $D = 64$ and $K = 32$ with predefined $J = 128$. 

For each sample  $X_{i}$, with $X_{i} = \{x_{1},...,x_{J}\}$, where $x_{j} \in \mathbb{N}^{0}$, for each $j = 1,..,J$, is  the number of behaviors in a given time window. The sample $X_{i}$ is transformed into a grayscale image $M_{i}$ of dimension $D\times D$  pixels. Thus,  $X_{i}$ is mapped to $M_{i}$, where $M_{i} =  \{m_{1},..,m_{J}\}$ and $m_{j}$ represents a variable, for each $j = 1,..,J$. Hence, $x_{j} = m_{j}$. We convert each tabular input data from decimal integer into binary format, denoting as $x_{j} = x^{\prime}_{j}$, so $ x^{\prime}_{j} = m_{j}$.

We extract each binary digit $p^{\prime}_{k}$ from $ x^{\prime}_{j} $ and get a set of 0 and 1 within the same order as the original binary digits, presenting as $ x^{\prime}_{j} = \{p^{\prime}_{1},...,p^{\prime}_{k}\}$, where $p^{\prime}_{k} \in \{0, 1\}$ and $k \leqslant K$. In other words, if $k > K$, it means $ x^{\prime}_{j}  > 2^{32} (4,294,967,296) $ as $K=32$ in proposed scenario. In such case, we set $x^{\prime}_{j} =  4,294,967,296$ and $p^{\prime}_{1} = p^{\prime}_{2} =... = p^{\prime}_{K} = 1$, because 4,294,967,296 is large enough to include majority cases in reality. Any larger number could be an outlier. We assign each $p^{\prime}_{k}$ to a pixel data $p_{k}$ on $m_{j}$, bringing us to $ m_{j} = \{p_{1},...,p_{k}\}$.  Deriving from decimal code RGB, we acquire $p_{k} = (255, 255, 255)$ (white) if $p^{\prime}_{k} = 0$ while $p_{k} = (0, 0, 0)$ (black) indicating $p^{\prime}_{k} = 1$. Additionally, $p_{k}$ on the image $M_{i}$ can be located by row number $r$ and column number $c$, for each $r,c = \{1,...,D\}$ due to the dimension of  $M_{i}$ is $D\times D$ pixels. By combining location and color property, we obtain $p^{r, c}_{k} = (0, 0, 0)\lor(255, 255, 255) $ for indicating a pixel data. We also use the same superscript $r$ and $c$ to denote the row and column of $M_{i}$ and $m_{j}$ so that we can get  $M^{r, c}_{i} = (0, 0, 0)\lor(255, 255, 255) $ and $m^{r,c}_{j} = (0, 0, 0)\lor(255, 255, 255) $ through inheriting the value of  $p^{r,c}_{k}$. 
\subsubsection{Mapping Algorithm}
After successfully converting each binary digit into a pixel format, mapping $r$ and $c$ among $M_{i}$, $m_{j}$ and $p_{k}$ is crucial. Basically, we design the PRS that  fits two variables $m_{j}$ in one line of pixels on image $M_{i}$. The order of input $\{p_{1},...,p_{k}\}$ for $m_{j}$ with odd $j$ is from left to right and for $m_{j}$ with even $j$ from right to left. The remaining algorithm explains the mapping procedure.  

\begin{figure}[b]
  \includegraphics[width=\linewidth]{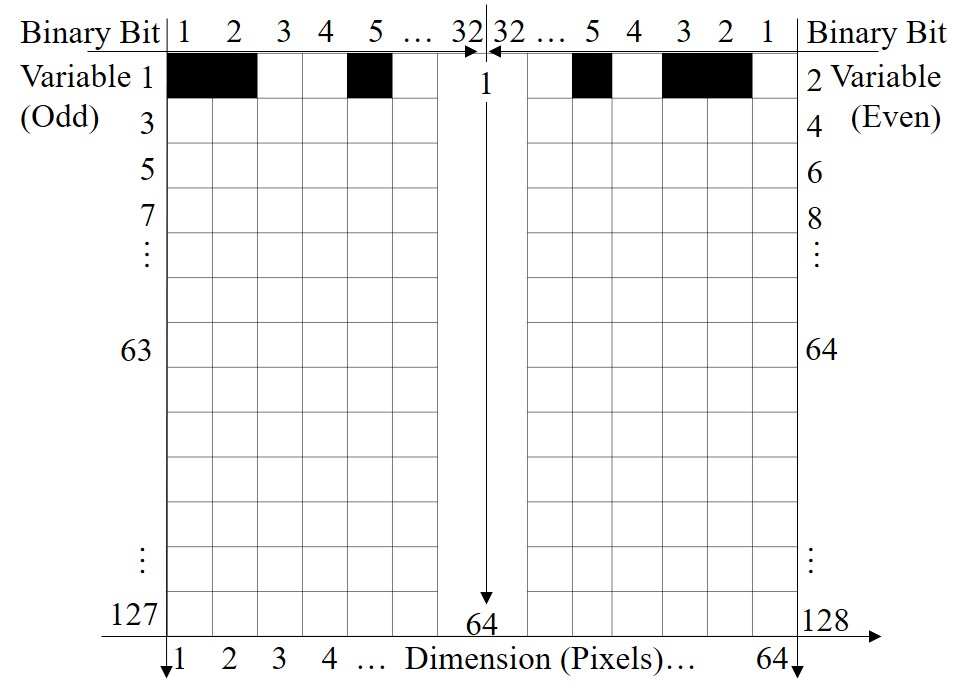}
  \caption{A Demonstration of PRS as $x_{1} = 19$ and $x_{2} = 22$ }
  \label{fig:PRS}
\end{figure}

\begin{itemize}[label={}, labelsep=0pt, leftmargin=0pt]
\item \textbf{Algorithm} \textit{PRS} $(M, m, p)$
\item \textbf{Input:} A transparent $D\times D$ pixels image $M_{i}$ containing information of $X_{i}$;  $m_{j}$: pictorial information of $x_{j}$; $\{p\}$: set of pixel data of  each $m_{j}$
\item \textbf{Output:} A sample image $M^{\prime}_{i}$ representing $X_{i}$
\item \textbf{Begin}
\begin{itemize}[label={},labelsep=0pt, leftmargin=5pt]
\item $J = 128, D = 64, K = 32, j = 1, j \leqslant J$
\item \textbf{for} $m_{j} \in \{M_{i}\}$
\item \hspace{0.5 cm} $r = (j \div 2) + 1$
\item \hspace{0.5 cm} \textbf{if} $j \bmod 2 = 1$ \textbf{then} 
\item \hspace{1 cm} $M^{r, (1,...,K)}_{i} = m_{j}$ \hspace{1 cm} \textit{(\# odd\#)}
\item \hspace{1 cm} $c = 1$
\item \hspace{1 cm} \textbf{for} $c \in [1,K]$
\item \hspace{1.5 cm} \textbf{if} $k \leqslant 0$ \textbf{then} 
\item \hspace{2 cm} $M^{r, c}_{i} = m^{r,c}_{j} = (255, 255, 255)$
\item \hspace{1.5 cm} \textbf{else} 
\item \hspace{2 cm} $M^{r, c}_{i} =  m^{r,c}_{j} = p^{r,c}_{k}$
\item \hspace{1.5 cm} \textbf{end if} 
\item \hspace{1.5 cm} $k - 1$
\item \hspace{1.5 cm} $c + 1$
\item \hspace{1 cm} \textbf{end for} 
\item \hspace{0.5 cm} \textbf{else}
\item \hspace{1cm} $M^{r, (D-K+1,...,D)}_{i} = m_{j}$ \hspace{1 cm} \textit{(\# even\#)}
\item \hspace{1 cm} $c = D$
\item \hspace{1 cm} \textbf{for} $c \in [D-K+1,D]$
\item \hspace{1.5 cm} \textbf{if} $k \leqslant 0$ \textbf{then} 
\item \hspace{2 cm} $M^{r, c}_{i} = m^{r,c}_{j} = (255, 255, 255)$
\item \hspace{1.5 cm} \textbf{else} 
\item \hspace{2 cm} $M^{r, c}_{i} =  m^{r,c}_{j} = p^{r,c}_{k}$
\item \hspace{1.5 cm} \textbf{end if} 
\item \hspace{1.5 cm} $k - 1$
\item \hspace{1.5 cm} $c - 1$
\item \hspace{1 cm} \textbf{end for} 
\item \hspace{0.5 cm} \textbf{end if}
\item \hspace{0.5 cm} $j +1 $
\item \textbf{end for}
\end{itemize}
\item \textbf{End}
\end{itemize}
Consider a sample $X_{1}$, and let 
\begin{center}
 $x_{1} = 19$ and $x_{2} = 22$ 
\begin{flushleft} So the binary transformation will be  \end{flushleft} 
$x^{\prime}_{1} = \{1,0,0,1,1\}$ and $x^{\prime}_{2} = \{1,0,1,1,0\}$.
\end{center}
 The mapping procedure will be as
\begin{IEEEeqnarray*}{LCL}
M^{1, (1,...,32)}_{1} &=& m^{1, (1,...,32)}_{1} \\
                                  &=& \{m^{1, (1,...,5)}_{1} , m^{1, (6,...,32)}_{1}\}\\
                                  &= & \{p^{1,5}_{1} = (0,0,0),\\
                                  &&p^{1,4}_{2} = (255,255,255),\\
                                  &&p^{1,3}_{3} =(255,255,255),\\
                                  &&p^{1,2}_{4} =(0,0,0),\\
                                  &&p^{1,1}_{5} =(0,0,0), \\
                                  && m^{1, (6,...,32)}_{1} = (255,255,255) \}, 
\end{IEEEeqnarray*}
\begin{IEEEeqnarray*}{LCL}                               
M^{1,(33,...64)}_{1} &=& m^{1, (33,...,64)}_{2} \\
                                     &=& \{m^{1,(33,...59)}_{2}, m^{1,(60,...64)}_{2}\} \\
                                     &=& \{m^{1,(33,...59)}_{2} = (255,255,255),\\
                                     &&p^{1,60}_{1} = (0,0,0),\\
                                     && p^{1,61}_{2} = (255,255,255),\\
                                     &&p^{1,62}_{3} = (0,0,0),\\
                                    && p^{1,63}_{4} = (0,0,0),\\
                                    &&p^{1,64}_{5} = (255,255,255)\}. 
\end{IEEEeqnarray*}
Put it together and generate the image 
\begin{center}
$M^{\prime 1, (1,...,64)}_{1} =\{M^{1, (1,...,32)}_{1} , M^{1, (33,...,64)}_{1}\} $
\end{center}
The remaining variables of $X_{1}$ should be mapped in a same fashion. Fig.~\ref{fig:PRS}  shows a partially generated image based on the described example.
\subsubsection{Padding}
In case the number of variables $J < 128$, we bring it up to the full size $J = 128$ and apply PRS as $D = 64$ and $K = 32$, automatically padding white pixels $(255,255,255)$ on each empty variable row. 

For $J>128$, there are four scenarios (S). 
\begin{enumerate}[label={\textbf{S\arabic*:}}]
\item $J \bmod 4 = 0$ \\
$\Rightarrow$ applying PRS with increased size of $D$ and $K$;
\item  $J \bmod 4 = 2$ \\
$\Rightarrow$ $D = J/2$ and $K = (J-2)/4$\\
$\Rightarrow$ setting $M^{(1,...,D), K+1}_{i} = (255, 255, 255)$ (padding on column $K+1$)\\
$\Rightarrow$  applying PRS with $D$ and $K$;
\item  $(J+1) \bmod 4 = 0$\\
$\Rightarrow$ $D = (J+1)/2$ and $K = (J+1)/4$\\
$\Rightarrow$ applying PRS with $D$ and $K$ (automatically padding);
\item  $(J+1) \bmod 4 = 2$\\
$\Rightarrow$ $D = (J+1)/2$ and $K = (J-1)/4$\\
$\Rightarrow$ padding $M^{(1,...,D), K+1}_{i} = (255, 255, 255)$\\
$\Rightarrow$ applying PRS with $D$ and $K$ (automatically padding);
\end{enumerate}
\subsection{Convolutional Neural Network}
Convolutional neural network (CNN) is defined as deep learning algorithms,  primarily using to solve difficult image-driven pattern recognition tasks \cite{o2015introduction}. CNNs are comprised of three types of layers, namely convolutional, pooling, and fully-connected layers. 

\begin{figure}[t]
  \includegraphics[width=\linewidth]{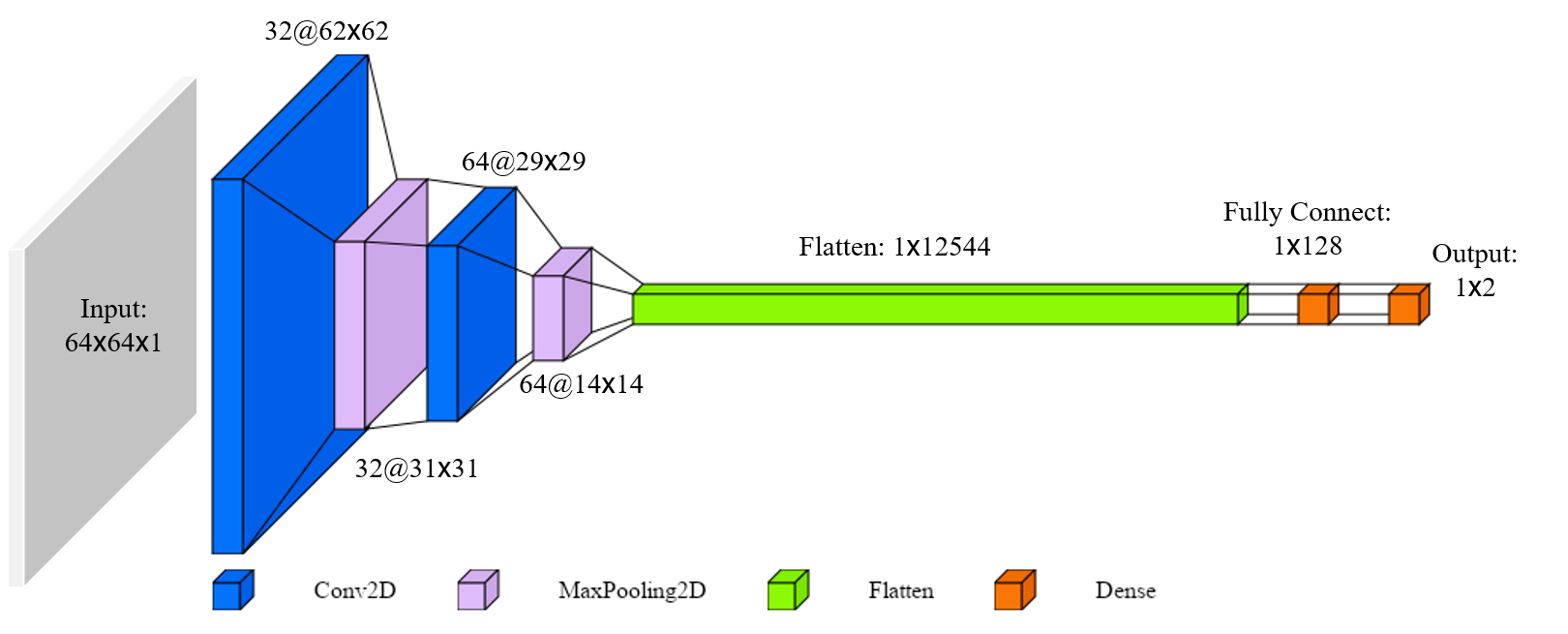}
  \caption{CNN Model Architecture}
  \label{fig:CNN}
\end{figure}

Convolutional layers are designed to learn procedure of feature representations of the inputs. In convolutional layers, a series of different sizes convolution kernels/ filters are implemented for computing different feature maps.  Mathematically, for each input $\mathbf{X}^{l}_{p,q}$ at location$(p,q)$ of the $l-th$ layer, the feature value $\mathbf{v}^{l}_{p,q,n}$ is calculated as
\begin{equation}
\mathbf{v}^{l}_{p,q,n} = \mathbf{w}^{l^T}_{n} \mathbf{X}^{l}_{p,q} + \mathbf{b}^{l}_{n} \label{eq2}
\end{equation}
where $n$ represents the $n-th$ feature map of the $l-th$ layer. $\mathbf{w}$ and $\mathbf{b}$ are the weight vector and bias term. Additionally $\mathbf{w}^{l}_{n}$  is shared with the next layer.
In our model, two two-dimensional (2D) convolutional layers are proposed. In the first hidden layer, the input image with the dimensions of 64 by 64 is convolved with 32, 3 by 3, filters. For the second layer, 64, 3 by 3, filters are applied to the inputs for capturing 64 feature maps.
The rectified linear activation function (ReLU) is used as nonlinear activation function on top of convolution in our model\cite{agarap2018deep}. The activation value $\mathbf{A}^{l}_{p,q,n}$ can be computed as:
 \begin{equation}
\mathbf{A}^{l}_{p,q,n} = \max( \mathbf{v}^{l}_{p,q,n}, 0)  \label{eq3}
\end{equation}
Over each convolutional layer, we operate a max pooling layer to reduce the dimensions through extracting maximum value from every 2 by 2 patch on input data\cite{gu2018recent}. Thus, after the second max pooling layer, 64, 14 by 14, feature maps stack together.  Then, a flatten layer is added on to shape pooled feature maps into a one-dimensional (1D) array (size 12544) of numbers (or vectors). A fully connected layer (known as dense layer) is imposed to connect each of 12544 inputs to each of 128 outputs. On this step, a ReLU activation function as described above is used. As a final touch, we add a softmax unit with two features, representing 0 and 1 classes. Fig.~\ref{fig:CNN} shows the architecture of our CNN model. 
\subsection{Conditional Generative Adversarial Network}
Reference \cite{goodfellow2014generative} first introduced Generative Adversarial Network (GAN) as a pair of neural networks:  a generative model G and a discriminative model D. In the G model,  input noise vectors $\boldsymbol{z}$ with a predefined probability distribution $p_{z}$ are mapped from the noise space $N_{z}$ to the image space $N_{x}$ for capturing the distribution $p_{\mathrm{data}}$ of all possible data in $N_{x}$, as D model is to distinguish the real data from $p_{\mathrm{data}}$ and fake ones synthesized by G model. Model G and D are simultaneously trained and optimized in an adversarial situation.  A value function is proposed by \cite{goodfellow2014generative} as :
 \begin{equation}
 \begin{split}
  &\min_{G} \max_{D} V (D, G)  = \\
   &\mathbb{E}_{\boldsymbol{x} \sim p_{\mathrm{data}} (\boldsymbol{x})} [ \log D (\boldsymbol{x}) ] + \mathbb{E} _{\boldsymbol{z} \sim p_{z} (\boldsymbol{z})}  [ \log (1 - D(G(\boldsymbol{z}))) ] \label{eq4}
 \end{split} 
\end{equation}
Conditional GAN (cGAN) \cite{mirza2014conditional, gauthier2014conditional} is an extension of GAN model with certain information (``conditions'') included in the inputs. In our model, conditions $\boldsymbol{y} \in N^y$ are class labels, 0 or 1, representing benign or malign. $\boldsymbol{y}$ is fed into both $G$ and $D$ models as additional input layer. Therefore, the new value function is developed as:
\begin{equation}
 \begin{split}
  &\min_{G} \max_{D} V (D, G)  = \\
   &\mathbb{E}_{\boldsymbol{x} \sim p_{\mathrm{data}} (\boldsymbol{x})} [ \log D(\boldsymbol{x}|\boldsymbol{y}) ] + \mathbb{E} _{\boldsymbol{z} \sim p_{z} (\boldsymbol{z})}  [ \log (1 - D(G(\boldsymbol{z}|\boldsymbol{y}))) ] \label{eq5}
 \end{split} 
\end{equation}
\begin{figure}[b]
  \includegraphics[width=\linewidth]{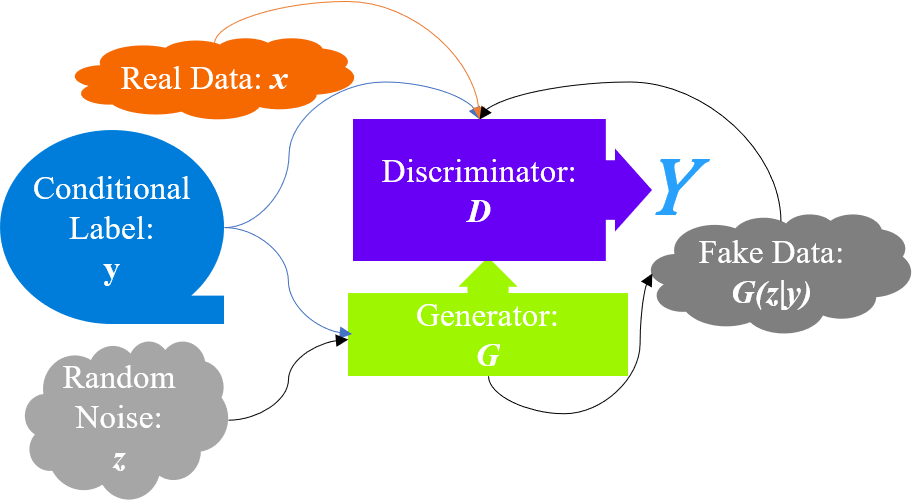}
  \caption{cGAN Model Demonstration}
  \label{fig:CGAN}
\end{figure}
As mentioned above, for both G and D model, we first need to create a layer embedded for class input and merged with image inputs. After the first step, D model is composed with 3 hidden layers with 128, 256, and 516 feature outputs, each followed by a Leaky version of ReLU activation layer (LeakyReLU) with threshold $\alpha = 0.2$, allowing a small gradient $0.2 * x$ when the unit $x < 0$. The added out layer is activated with a sigmoid function. Meanwhile, G model has 3 hidden layers with feature outputs 256, 512 and 1024, followed by same LeakyReLU. An batch normalization layer is also applied after each hidden layer in G model. The out layer is activated by tanh function.
\section{Experiments}
\subsection{Pictorial Transformation}
\begin{figure}[t]
  \includegraphics[width=\linewidth]{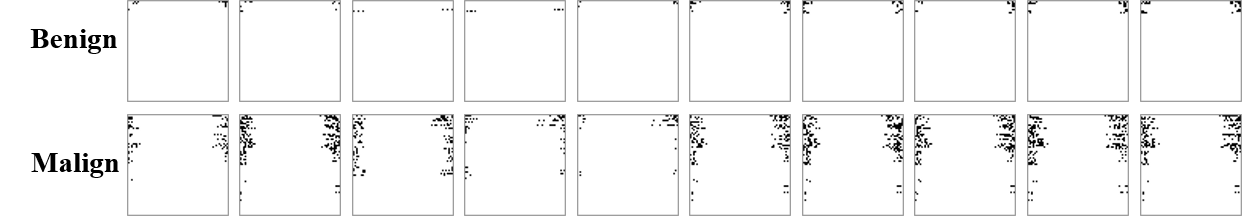}
  \caption{Benign and Malign Samples Generated through PRS}
  \label{fig:Samples}
\end{figure}

Training and testing data are processed and transformed into pictorial format through the PRS system.  Samples with two classes are presented on Fig.~\ref{fig:Samples}.  
Comparing to samples with malign label, images with benign label have less amount of black colors, most of which are only scattered around corners and edges, indicating variables with smaller values. This result matches our expectations. According to observations on tabular data, we notice that large numbers are concentrated in malign samples. Additionally, a larger proportion of non-zero variables per sample appears frequently in malign class. Therefore, on average, more black pixels occur on malign samples and are inclined to drift from both left and right sides to the middle part of the image.  
\subsection{Generating Artificial Malware Images}
\begin{figure}[ht]
  \includegraphics[width=\linewidth]{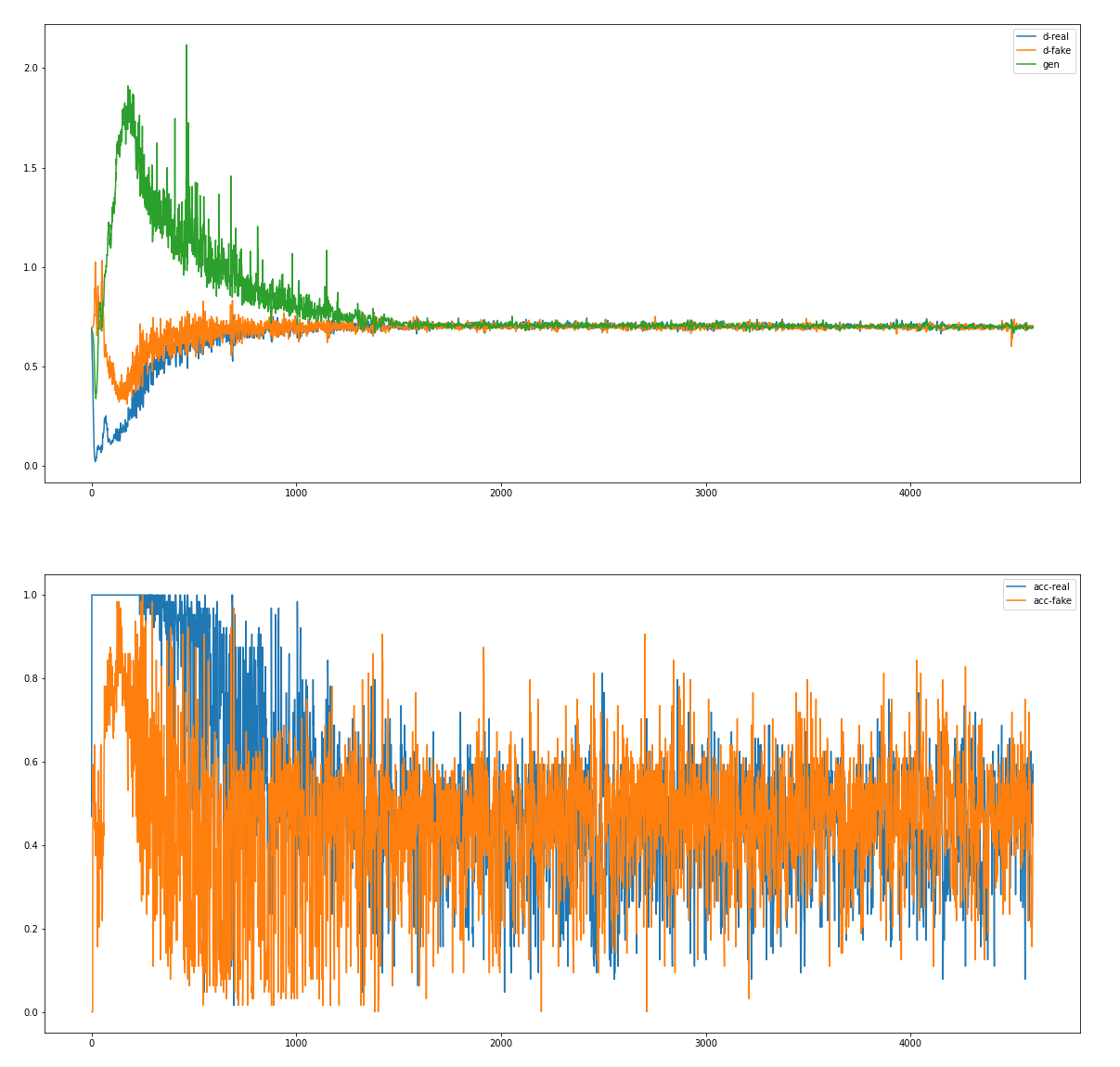}
  \caption{The Learning Curve of cGAN Model.
  The top panel shows  the discriminator loss for real images (blue), discriminator loss for generated fake images (orange), and the generator loss for generated fake images (green). The bottom panel shows  the discriminator accuracy on real (blue) and fake (orange) images during training. On both panels the convergence is achieved at around 1500 iterations.}
  \label{fig:accloss}
\end{figure}

As introduced in previous section, 3000 benign samples and 3000 malign samples are fed to the cGan model. We experiment with different options and combinations of hyperparameters to tune the model. Finally, we select 0.0002 learning rate for the optimizer of D model. The best performance cGAN model is realized with $epochs = 100$ and $batch\_ size = 128$. Fig.~\ref{fig:accloss} shows the learning curve of cGan model. We can conclude that after passing initial somewhat erratic stage, models D and G reach convergence after about 1500 iterations. At convergence, the loss of G model and the loss of D model on both discriminating real and generated data are all stabilized around $0.69$, meaning that for both models, improvements to one model will not come at the expense of the other model.  A point of equilibrium between the two competing concerns is reached. 

\begin{figure}[!h]
  \includegraphics[width=\linewidth]{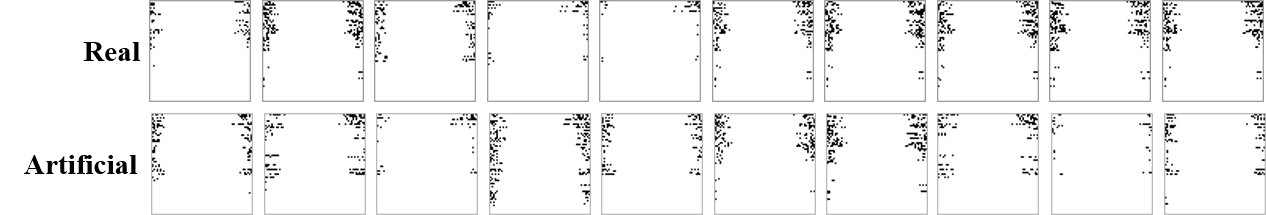}
  \caption{Artificial Malign Samples Generated by cGAN Model}
  \label{fig:gansample}
\end{figure}

Samples of malware images generated by the model are presented in Fig.~\ref{fig:gansample}. Artificial malware image samples exhibit very similar patterns with the real ones. In both cases, a large amount of black pixel blocks are dispersed on the images. This procedure brings us 1535 artificial malign samples, which will be included in the training data for the next step. 

\subsection{Comparing Malware Detection Performances}
Two sets of training data have been obtained, namely, the original dataset and the one containing artificial samples generated through cGAN model. We feed both sets to CNN models and  compare their performances on detecting malware from the same testing set. The goal of the comparison involves finding whether visualized malware according to PRS could be easily identified by similar models without complex feature tuning procedures. 

\begin{figure}[t]
  \includegraphics[width=\linewidth]{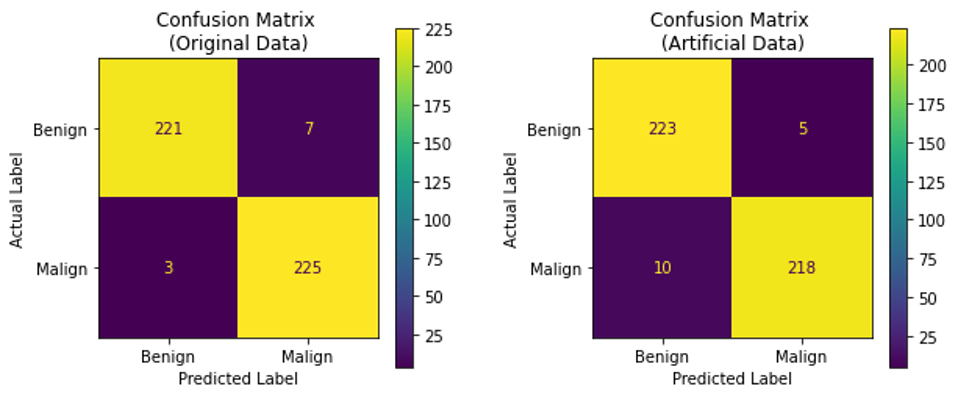}
  \caption{Confusion Matrix for Two Scenarios}
  \label{fig:cof}
\end{figure}

The baseline model proposed in the previous section is composed according to the original dataset.  We also implement 5-fold cross validation technique on the training set for evaluating the model. The average accuracy of the baseline model during cross validations is $98.51\%$. We apply the same procedure on the dataset with artificial images, bringing us a slightly lower average validation accuracy, $97.26\%$.  We proceed to fit the models to the same testing dataset in order to compare detection performances using  different training sets. We summarize the performance of both CNN models in Table ~\ref{tab:clsf} and Fig.~\ref{fig:cof}.

\begin{table}[tp]
 \begin{threeparttable}
  \centering
  \caption{Classification Reports of CNN Models}
   \label{tab:clsf}
    \begin{tabular*}{\linewidth}{@{\extracolsep{\fill}}ccccccc@{}}
    \toprule
    \multicolumn{1}{c}{} &
      \multicolumn{2}{c}{Precision} &
      \multicolumn{2}{c}{Recall} &
      \multicolumn{2}{c}{F1-score} 
      \\
    \multicolumn{1}{c}{} &
      A &
      B &
      A &
      B &
      A &
      B 
      \\
    \midrule
    \multicolumn{1}{c}{Benign} &
      0.9866 &
      0.9571 &
      0.9693 &
      0.9781 &
      0.9779 &
      0.9675 
      \\
    \midrule
    \multicolumn{1}{c}{Malign} &
      0.9698 &
      0.9776 &
      0.9868 &
      0.9561 &
      0.9783 &
      0.9667 
      \\
    \midrule
    \multicolumn{1}{c}{Acc} &
       &
       &
       &
       &
      0.9781 &
      0.9671 
      \\
    \midrule
    macro avg &
      0.9782 &
      0.9673 &
      0.9781 &
      0.9671 &
      0.9781 &
      0.9671 
      \\
    \midrule
    weighted avg &
      0.9782 &
      0.9673 &
      0.9781 &
      0.9671 &
      0.9781 &
      0.9671 
      \\
    \bottomrule 
    \end{tabular*}%
    \begin{tablenotes}\footnotesize
    \item[*] A: results of baseline model training on the (original) dataset without artificially generated images; 
    \item[*] B: results of model training on the dataset with artificially generated images
    \item[*] Support: 228 benign samples and 228 malign samples for both A and B
    \end{tablenotes}
    \end{threeparttable}
\end{table}%

According to F1-score on Table ~\ref{tab:clsf}, we can see that model trained on the original dataset shows slightly better performance than the one trained on the dataset with artificial images. The accuracy score is $1.1$ percentage higher. As reported in the confusion matrix in Fig.~\ref{fig:cof}, it is noticeable that the model performing on the artificial dataset generates fewer false positives and has better recognition efficiency of malign samples.  In other words,  when the model predicts malware based on the dataset with artificial images, it is correct $97.76\%$ of the time. In contrast, with the original dataset, it is correct $96.98\%$ of the time. However, the baseline model is more sensitive and able to correctly identify $98.68\%$ of all malign samples, about $3\%$ higher than the second one. Thus,  the malware detection performances are comparable. 

We explore several possible reasons for observed minor differences between the two models. First, information loss could occur during the synthesizing procedure of cGAN. Although under the name of ``grayscale'', original images generated through the PRS have two colors, black and white. In cGAN, the implementation of 1-color channel is not sufficient to filter various gray colors, leading to unnecessary colors synthesized on the image. Eliminating gray colors in cGAN would likely to increase the quality of the generated images. Second, the detection model, CNN, is based on the original dataset. A more aggressive tuning procedure could be employed in order to guarantee a higher accuracy of the synthetic dataset. Finally, it is worthy to consider improving CNN model for classifying this specific type of images. For example, our preliminary experiments without max pooling layers obtained slightly higher average accuracy, even though the learning curve still remained unstable after certain number of iterations. Additionally, alternative neural net architectures, such as Vgg16 \cite{simonyan2014very}, AlexNet \cite{krizhevsky2012imagenet}, and others could be beneficial at expense of the increased computational costs.

\section{Conclusion and Future Work}
In this work, we have proposed a Pictorial Representation System (PRS) of transforming tabular data into images, visualizing the behavioral patterns on malware for identifying purpose. We have developed a conditional Generative Adversarial Network (cGAN), by which the same type of malware images could be successfully synthesized to tackle imbalanced data problems caused by uneven distribution of benign and malicious software. We have implemented Convolution Neural Networks (CNNs) to compare the classification accuracy with and without training on synthetic images. We obtained comparable malware detection performance on both datasets. 

The proposed PRS approach can be used to preserve malware identity not only for being easily recognized by human eyes but also for analyzing malicious behavioral patterns. Our procedure involved with CNN and cGAN can be used to achieve synthetic augmentation of malware datasets as well as for improving the robustness of malware detection solutions.  Moreover, we have introduced a padding method to extend the PRS for generating images of arbitrary sizes. In the future, we plan to explore other methods for extensions of the PRS and combine improved CNN and cGAN in a wide range of applications for malware detection purpose. 

\bibliographystyle{IEEEtran}
\bibliography{paper}

\end{document}